\documentclass[preprint]{aastex}
\usepackage{emulateapj5}
\usepackage{epsfig}
 
\begin{document}

\title{Prospecting for spiral structure in the flocculent outer Milky Way Disk 
with color magnitude star counts from the 2 Micron All Sky Survey}

\author{A.~C.\ Quillen
}
\affil{Department of Physics and Astronomy, University of Rochester, Rochester,
NY 14627; aquillen@pas.rochester.edu}

\begin{abstract}  
Using star counts in both color and magnitude from the 
Two Micron All Sky Survey (2MASS) 
Second Incremental Release Point Source Catalog
we search for evidence of non-uniform extinction and stellar population
density changes in the Galactic Plane.
Extinction causes the entire main sequence to shift
toward redder colors on a color magnitude diagram.
A local increase in the stellar density causes 
an increase in the star counts along a line parallel to 
the main sequence.
We find streaks in star count color magnitude contour plots 
along the angle of the main sequence
which are likely to be caused by distant gas 
clouds and stellar density variations.
The distance of a gas cloud or stellar density change
can be estimated from the location of the shift in the star count contours.
We identify features in these diagrams which are coherent 
across at least 10 degrees in Galactic longitude.  
A series of features is seen at the plausible distance of the expected 
Perseus spiral arm
at a distance of 2 to 4 kpc from the sun.  However other features as
prominent are found at both at larger and smaller distances.
These structures are over 300 pc in size
and so likely to be associated with large scale coherent structures
in the gas distribution such as weak spiral arms. 
The presence of multiple and weak spiral arms, and lack
of strong ones suggests that the outer 
Milky Way is flocculent in its morphology.
\end{abstract}  

 
\section{Introduction}

It has long been recognized 
that star counts or the
number of stars per magnitude bin per unit area on the
sky are strongly affected by extinction, or the
absorption of light by dust  (e.g., \citealt{seeliger,mccuskey}).
Because extinction is particularly strong in the plane
of the Milky Way, it has not been possible to 
study the spiral structure of our Galaxy deep within the Galactic Plane
using star counts. 
However, above the scale height of the gas and dust,
at Galactic latitudes outside the ``Zone of avoidance''
or greater than $10^\circ$ from
the Galactic Plane, it is possible to measure
the distribution of stars in the galaxy. 
For example star counts from the Sloan Digital Sky Survey well above 
the Galactic Plane 
have recently been used to measure 
the scale heights of the thin and thick stellar disks
\citep{chen} and uncover structure in the halo \citep{yanny}.

Because of the sensitivity to extinction, rather
than probe for galactic structure, star counts
are often used to measure extinction,
as first illustrated with the Wolf diagram, \citealt{wolf,trumpler},
or make extinction maps 
in the visible bands (e.g., \citealt{dickman}) and
at very high levels of extinction in molecular clouds
in the near-infrared (e.g., \citealt{lada,cambresy,lombardi}).

The 2MASS team has recently released 
a significant fraction of the sky at a depth of about 15 mag in $K_s$ band
(the Second Incremental Release).
For information on the 2MASS Second Incremental Release Point Source Catalog 
see http://irsa.ipac.caltech.edu.
Because 2MASS is carried
out between 1 and 2 microns, the effect of extinction is greatly reduced
compared to that at visible wavelengths.
2MASS therefore provides us with a unique opportunity to probe
for structure within the plane of our Galaxy.

\section{Number Counts in a color-magnitude diagram}

To detect structure in the halo
\cite{yanny} selected stars in a narrow
range of color.  On the main sequence color effectively determines
the luminosity of the star and so its distance can be determined
from the observed magnitude.
While this technique worked well to uncover structure
in the halo, we now explain why it presents difficulties 
when used to search for structure in the Galactic Plane 
using 2MASS data.

A difference in $J-H$ of 0.1 results in a change
in absolute magnitude on the main sequence of about a magnitude
(see Fig.~1)
and this would correspond to a factor in distance of 1.6.   
The nearest spiral arms are about 2 kpc 
away from the sun (e.g., \citealt{vallee}) 
and may only have thicknesses of a few hundred pcs,
so to see an enhancement in the star counts
we can only tolerate distance uncertainties that are 
less than factors of 1.1.  This corresponds to a magnitude uncertainty
of less than 0.2 which is difficult to achieve
using an infrared color section for stars at the distance where
we expect spiral arms.
We expect that magnitude uncertainties resulting from
the strong dependence of absolute magnitude on color
of the main sequence
will wipe out any evidence for spiral structure in a given
color selected sample.
Furthermore because of reddening  associated
with extinction, a color selection will not
restrict the stars to a give spectral type or absolute magnitude.
So despite the fact that this approach was successful
for detecting structure in the halo, if we only
search in one color bin it is unlikely 
to work in the Milky Way disk using 2MASS data.

So how do we get around this problem?
We must use information from stars which cover a range of 
observed colors.
The main sequence provides us with a useful 
relation between color and absolute magnitude.
A change in stellar population caused by recent star formation
causes a local increase in the number of blue stars, whereas
extinction will cause the entire main sequence to shift
toward redder colors on a color magnitude diagram.
If we see features on a color magnitude number count
diagram (number counts per unit color and magnitude bin) 
that extend along the direction of the main sequence, then
we expect that those features are caused by local
changes in either extinction or the local stellar population density.

In Figure 1 we show star counts from a 2.5 square
degree field taken from the 
2MASS Second Incremental Release Point Source Catalog.
Stars were counted in an array of bins defined by both
color and magnitude.  Magnitude bins were 0.1 mag wide
in either $J$ or $K_s$ bands
and $J-H$ color bins were 0.025 mag wide.   
To aid the contouring algorithm we  slightly
smoothed the array with a smoothing function 3 pixels
high and wide that is shaped like a pyramid with half the flux
in the central pixel.
The minimum contour shown contains 5 stars per bin
and we show a total of 12 contours spaced logarithmically with the 
maximum being the maximum value of counts in the array.
For the 2MASS Point Source Catalog the
Level 1 requirements were to achieve a signal to noise
of S/N$= 10$ at 15.8, 15.1, and  14.3 mag
in $J, H$ and $K_s$ bands respectively.
The actual achieved limiting magnitudes  at this signal to noise 
are about 0.5 magnitude better than expected.
Typical photometric errors are about 0.03 mag for 13,12.5 and 12.5
magnitudes in $J, H$ and $K_s$ band respectively
and about 0.1 mag for 16.3,15.5 and 15.0 mag respectively.

As a stellar population becomes more distant, the luminosity
function is pushed to higher magnitudes on this figure.  However
extinction both reddens and decreases the luminosity and so
affects the entire main sequence along the same vector
in the diagram.
A region containing a young population increases
the number of blue stars at a particular magnitude range or distance.
An increase in the stellar density at a particular distance
will cause an increase in counts again along an angle
parallel to that of the main sequence.  This will
cause a shift in the contours toward the left.
In Figure 1 we have over-plotted the main sequence at a solar
metallicity and distance of 1kpc for 3 different aged populations.
Diagonal streaks in the contour levels are observed
which are along the same angle as the main sequence but offset
in magnitude.  
Because we see bends in the contours that extend over 
a magnitude in J and K and 0.2 mag in color, they are likely
to be real and not to due Poisson noise.
These diagonal streaks are probably due to 
extinction at a distance exceeding the main sequence model points.
The distance to the gas cloud responsible for the extinction
can be estimated from the magnitude shift of the main sequence.
A streak in the contours is seen in both
$J$ and $K_s$ vs $J-H$ band 
color magnitude plots corresponding
to an extended enhanced region of extinction or gas cloud at a distance 
of about 2.5 kpc from the sun.

\section{Prospecting for Spiral Structure}

We now look for coherent structures seen in the color magnitude diagrams 
in different locations in the Galactic Plane.
There are three regions in the Second Incremental release
that contain the Galactic Plane and not the Galactic Center. 
These regions have
Galactic longitudes which range 
$glon=220$--$250^\circ$, 50--$80^\circ$ and 170--$190^\circ$.
All three regions should contain the 
the Perseus Spiral arm (see Fig.~2), however in
the 50-80$^\circ$ region the arm is 
distant at 5-7 kpc and so it should
be impossible to detect it using main sequence stars.
Furthermore because the spiral arms increase with distance
with increasing Galactic longitude, even weak spiral
arms in the region between the Carina and Perseus arms at $glon \sim 80^\circ$
should not extend over a large range in galactic longitude. 
In the Galactic anti-center region between $glon= 170$---$190^\circ$ 
the Perseus spiral arm
should be roughly equidistant from the sun at a distance of 2 kpc 
but little extinction is evident in this direction.
So we begin by concentrating on the region between 220 and $250^\circ$.

In Figure 3 we show color magnitude number count diagrams in regions
ranging from 220-250$^\circ$ with Galactic latitudes
$glat = 0$ to $-2$ or $-3$ to $-1^\circ$.
In a few regions pieces of the sky were not covered by the data release
so the counts were corrected by dividing by the percent of sky area covered.
The $K_s$ band plot corresponding to the $J$-band plot in Fig.~3a
is not shown since it is quite similar to Fig.~3a.
In this region between we see streaks
(denoted by a dotted line in Fig.~2)
that begin at $J\sim 14.5$ at $glon=220$ and end 
at $J \sim 15.5$ at $glon \sim 235$.
This feature is also seen in the $K_s$ color mag contour plot (Fig.~3c).  
The location of the streaks is at somewhat larger magnitudes
for $glat =-3$ to $-1$ than for $glat = 0$ -- $2^\circ$
suggesting that there is a larger smooth extinction gradient
with radius at this Galactic latitude than at $glat=1^\circ$.
From the location of the feature in the $K_s$-band plots and how it
is offset from the main sequence at 1 kpc (see Figure 1) 
we can estimate its
distance to be $\sim 2.5$ kpc at $glon = 220$ extending
to $\sim 4$ kpc at $glon=235$.   This corresponds to the expected
location of the Perseus spiral arm (see Fig.~2) as extended from
our knowledge in existing spiral tracers \citep{vallee}.

We also see evidence for additional structures which
have not been identified by \cite{drimmel} and HI surveys.
See for example the
indent in contours at $glon=225$ to $235$ at $K_s\sim 13.8$ to $15$
which would correspond to extinction at a distance of
2--4 kpc but inside the Perseus arm which is expected to be
at a distance of 5 kpc at $glon=235^\circ$. 

Structure seen in these star count color magnitude
contour plots appears to be is correlated and extended.  
A streak in the contours
at one galactic longitude is likely to be present
in the diagram at lower and higher longitude, 
though systematically shifted toward more higher
magnitudes at higher galactic longitudes.  This makes
sense if the density perturbations both
in the gas and stars are spiral in shaped, or increasing
with distance with increasing Galactic longitude (see Fig.~2,4).

The structure see in these star count diagrams cannot
be solely due to extinction because local increases in the star counts
at higher magnitudes is also seen.  This implies that there
are also density variations as a function of distance from the sun.   
Again this is not
unexpected if we remember what other galaxies look
like (e.g., see Fig. 4 which displays a B-band image of M100.)

Though see strong evidence for local density and extinction
variations in the region around $glon \sim 230$, we do not features
associated with 
extremely strong spiral structure, neither evidence for
large variations in the stellar density or local extinction.
The most likely explanation for the structure that we do see is that
the Milky Way is flocculent in its morphology outside the solar
circle.

For a spiral arm outside the solar circle
we expect to encounter first an increase in extinction and then
an increase in the stellar density as a function of distance 
from the sun as a spiral arm is crossed (see Fig.~4).  
The affect of the extinction is therefore decreased by the 
increased density following the spiral arm.
Stellar density and extinction variations associated
with spiral structure should be easier to uncover 
using star counts in the Carina arm within the solar circle
because we expect that blue stars should be nearer than
the belt of extinction associated with the spiral arm.
This should emphasize features seen in the color magnitude star count plots,
rather than decrease them as is true when we look toward larger
Galactocentric radii.
Features associated with these spiral arms may also be easier
to detect because these spiral features should have 
higher gas densities, hence
heavier extinction, and stronger stellar density contrasts.

In Figure 5 we show color magnitude star count plots
taken from a strip along $glat = 2$-$3^\circ$ and between
$glon= 65$--$85^\circ$.
There is a lot of extinction at $glon\sim8^\circ$ which corresponds
to the Dark Nebula LDN-906 (Lynds 1962) which has an area of $15^\circ$
and must be quite nearby.
A number of features are seen in these plots which are probably
real because they extend over a fair range on an individual
plot.  However we do not find a high level of coherence between
the plots.  This is not surprising since at this
Galactic longitude we are looking between the Carina and Perseus
arms and the opening angle of the arms is such that elongated
gas clouds will not extend over a large range in Galactic longitude.
The large number of features seen in these plots is consistent
with a patchy ISM and the 
possibility that the outer Milky Way is flocculent in its morphology.

In Figure 6 we show color magnitude star count plots
taken from strips along $glat = -1$ to $1^\circ$, $-3$ to $-1^\circ$ 
and between $glon= 170$--$190^\circ$.  At  $glon\sim 170$ to $180^\circ$
we see evidence for extinction at a distance of about 2 kpc
(streaks beginning at $K_s \sim 13.8$).
This does corresponds to the expected location of the Perseus spiral arm, 
however the lack of any features at $glon \sim 190$ suggests
that it is not continuous.    There are also extended features further
out at $K_s \sim 14.5$ from $glon \sim 180$ to $190^\circ$.
We see no dominated  feature at expected location of the Perseus
arm but instead a variety of shorter features at both shorter 
and longer distances.

\section{Summary and Discussion}

In this paper we have presented color magnitude number count contour
plots constructed from the 2MASS Point Source Catalog in the Galactic
Plane. 
We demonstrate how streaks observed in these diagrams can
be used to estimate the distance of local regions of high
extinction.  A coherent structure is seen between Galactic
longitude 220 and $237^\circ$ which
corresponds to the probably location of the extension of the Perseus spiral arm.
Another coherent structure is seen with a steeper pitch
angle between Galactic longitude 230-240$^\circ$
at a distance of 2-4.0 kpc and may be a weaker spiral
feature.  Neither feature is particularly prominent.
In the region $glon\sim 170$ to $190^\circ$ we see a feature
at the probably location of the Perseus spiral arm but also 
also other coherent features at both larger and smaller distances.
Features seen in the star count contour plots are likely to be real
because they cover a significant range in color and magnitude
in each color magnitude star count plot and because nearby
lines of sight in the galaxy exhibit similar features.
At a distance of 2kpc a structure which extends over 10 degrees
has a length of 350 pc so the coherent structures picked
out from the star counts are likely to associated with spiral arms.  
The lack of strong spiral features and the evidence
for additional weak features suggests that the outer part
of Galaxy is flocculent in its morphology.

We suggest that color magnitude number counts could be used to
estimate the distances to large scale gas clouds at distances
up to a few kpc using blue main sequence stars.
It is possible that a similar technique could also reveal
structure out to much larger distances using giant stars.
Giant stars have the disadvantage that 
it is not straightforward to estimate the luminosity of
the star given its color but the advantage that they can
be seen much further away than blue main sequence stars.

Using star counts to measure
distances to atomic and molecular clouds is potentially a very powerful technique.
However to do this reliably we would need to integrate the stellar 
luminosity function as a function of distance from us and
devise techniques to differentiate between
patchy extinction, stellar density variations, mean extinction
variations and variations in the mean age of the stellar population.
All of these affects should affect the integrated number counts.
By combining information available in CO, HI  and far-IR images
of the Milky Way with 2MASS it may be possible to make
a map of the Milky Way with the detail of that shown
in Figure 4 of M100 rather than that currently known which
is represented in Figure 2.

\acknowledgments

This work could not have been carried out without helpful discussions
with Larry Helfer, Rob Selkowitz, Joel Green, Dan Watson, and Eric Blackman.

This research has made use of the NASA/IPAC Infrared Science Archive, 
and the NASA/IPAC Extragalactic Database (NED) which are
operated by the Jet Propulsion Laboratory, California Institute of Technology, under
contract with the National Aeronautics and Space Administration.
This publication makes use of data products from the Two Micron All Sky Survey, 
which is a joint project of the University of
Massachusetts and the Infrared Processing and 
Analysis Center/California Institute of Technology, funded by the National
Aeronautics and Space Administration and the National Science Foundation.

{}

\end{document}